\documentclass[a4paper,twocolumn,english,showpacs,nofootinbib,nobibnotes,aps,pra,floatfix,superscriptaddress]{revtex4-2}
\usepackage[utf8]{inputenc}
\usepackage{amsmath}
\usepackage{amsthm}
\usepackage{amssymb}
\usepackage{graphicx}
\usepackage[caption=false]{subfig}
\usepackage{xparse}
\usepackage{xcolor}

\usepackage{times}

\makeatletter

\let\oldproof\proof
\renewcommand{\proof}{\color{darkgray} \oldproof }
\newcommand{\trace}{\operatorname{Tr}}

\newcommand{\leakEC}{\operatorname{leak}_\text{EC}}

\newcommand{\one}{\mathds{1}}
\newcommand{\bra}[1]{\langle #1|}
\newcommand{\ket}[1]{|#1\rangle}

\newcommand{\ketbraa}[2]{| #1\rangle \langle #2|}
\newcommand{\ketbra}[1]{| #1\rangle \langle #1|}

\usepackage{braket}
\usepackage{dsfont}

\makeatother

\usepackage{babel}

\global\long\def\trace{\operatorname{Tr}}
\global\long\def\ketbra#1{\ket{#1}\!\bra{#1}}
\global\long\def\ketbraa#1#2{\ket{#1}\!\bra{#2}}

\global\long\def\one{\mathds{1}}

\newcommand{\kommentar}[1]{}

\NewDocumentCommand\opti{smmm>{\SplitList{;}}m} {
\begingroup%
\setlength{\belowdisplayskip}{-0.6\baselineskip}%
\IfBooleanTF{#1}{%
    \begin{alignat*}{2}
        & \underset{#3}{\text{#2}} & & #4 \\
        & \text{subject to~~}
        \ProcessList{#5}{ \insertopticonst }
        & &
    \end{alignat*}%
    }{%
    \begin{alignat}{2}
        & \underset{#3}{\text{#2}} & & #4 \\
        & \text{subject to~~}
        \ProcessList{#5}{ \insertopticonst }
        & & \nonumber
    \end{alignat}%
    }%
\endgroup%
}%
\newcommand\insertopticonst[1]{& & #1\\&}
\usepackage[normalem]{ulem}

\begin{document}

\title{High-dimensional quantum key distribution rates for multiple measurement bases}

\author{Nikolai Wyderka}
\affiliation{Institut für Theoretische Physik III, Heinrich-Heine-Universität Düsseldorf, Universitätsstr. 1, D-40225 Düsseldorf, Germany}

\author{Giovanni Chesi}
\affiliation{QUIT Group, Dipartimento di Fisica, Università degli Studi di Pavia, Via Agostino Bassi 6, 27100 Pavia, Italy}

\author{Hermann Kampermann}
\affiliation{Institut für Theoretische Physik III, Heinrich-Heine-Universität Düsseldorf, Universitätsstr. 1, D-40225 Düsseldorf, Germany}

\author{Chiara Macchiavello}
\affiliation{QUIT Group, Dipartimento di Fisica, Università degli Studi di Pavia, Via Agostino Bassi 6, 27100 Pavia, Italy}

\author{Dagmar Bruß}
\affiliation{Institut für Theoretische Physik III, Heinrich-Heine-Universität Düsseldorf, Universitätsstr. 1, D-40225 Düsseldorf, Germany}

\date{\today}

\begin{abstract}
We investigate the advantages of high-dimensional encoding for a quantum key distribution protocol. In particular, we address a BBM92-like protocol where the dimension of the systems can be larger than two and more than two mutually unbiased bases (MUBs) can be employed. Indeed, it is known that, for a system whose dimension $d$ is a prime or the power of a prime, up to $d+1$ MUBs can be found. We derive an analytic expression for the asymptotic key rate when $d+1$ MUBs are exploited and show the effects of using different numbers of MUBs on the performance of the protocol. Then, we move to the non-asymptotic case and optimize the finite key rate against collective and coherent attacks for generic dimension of the systems and all possible numbers of MUBs. In the finite-key scenario, we find that, if the number of rounds is small enough, the highest key rate is obtained by exploiting three MUBs, instead of $d+1$ as one may expect.
\end{abstract}
\maketitle
\section{Introduction}
The experimental realization of high-dimensional (HD) quantum systems has been boosting the implementation of large-alphabet communication protocols, which can outperform the standard ones based on two-level systems. This is the case, in particular, of quantum key distribution (QKD) protocols.
\\
The idea of exploiting quantum systems for cryptography started with the seminal work by Bennett and Brassard \cite{bb84}, 
who used two-dimensional states. This was later on also formulated in an entanglement based version \cite{bennett1992} (BBM92). Then, the search for more secure and robust protocols led to consider larger Hilbert spaces for the employed physical systems.
Nowadays, there are many feasible physical implementations of HD states with applications to QKD, such as spatial encodings \cite{walborn2006,Lib2024}, time-bins \cite{ali-khan2007,islam2017,nussbaum2015}, temporal modes \cite{brecht2015}, angular momenta \cite{mirhosseini2015}, polarization for biphoton fields \cite{Sekga2023} and path encodings \cite{canas2017,Zahidy2024}. 

Basically, the advantage that QKD takes from HD systems is twofold. On the one hand, the scaling of secret key rates with the logarithm of the dimension of the system \cite{cerf2002} allows to improve the performance of the protocols \cite{bruss2002}. On the other hand, it is possible to enhance the resilience to errors and/or the secure key rate by exploiting a larger number of mutually unbiased bases (MUBs) \cite{wootters1988}. Indeed, it is known that for every system it is possible to find at least three complementary bases and, if the dimension $d$ of the Hilbert space is the power of a prime, there are $d+1$ of them \cite{wootters1988,durt2010}. The first evidence that exploiting more than two MUBs improves the secure key rate of the protocol was given in Refs.~\cite{bruss1998,gisin1999}, where the well-known six-state protocol was investigated. The protocol generalizes the BB84 protocol by using three complementary bases instead of two. Then, in Ref.~\cite{cerf2002}, an upper bound on the error rate was derived in the presence of $d+1$ MUBs. In Ref.~\cite{bradler2016finite}, analytic expressions for the asymptotic key rates are derived for generic dimension and three MUBs. Moreover, a finite key rate analysis is developed in the same scenario, proving an enhancement in security against collective attacks. Much work has been done to numerically improve secure key rates through semi-definite programs \cite{winick2018,doda2021,brown2021,Hu2022,araujo2023}. In addition, a flexible analytic framework for the calculation of HD asymptotic key rates was developed in Ref.~\cite{huber2024}. From the experimental side, recently a scalable implementation of $d+1$ MUBs was proposed and realized with time bins in prime power dimensions \cite{ikuta2022}, showing that a protocol exploiting $d+1$ MUBs is doable and improves the robustness of QKD protocols.

Motivated by these results, we address a BBM92-like scheme \cite{bennett1992}. We analyse the generalized version, where the dimension of the Hilbert space is arbitrary and every allowed number of MUBs can be selected. 
In particular, we investigate the impact of the number of MUBs employed on the key rate and study the role of the dimension.
We derive the analytic expression of the asymptotic secret key rate when $d+1$ MUBs are used and provide numerical results for other cases.
In the non-asymptotic regime, we find upper bounds on key rates secure against collective and coherent attacks. In the case of collective attacks, we compare the optimizations of rates obtained from an uncertainty relation for smooth entropies \cite{renner2005smoothent,tomamichel2010,tomamichel2011uncertainty,tomamichel2012tight,grasselli2020quantum} with the ones obtained from the asymptotic equipartition property (AEP) \cite{renner2008security,grasselli2020quantum}. The result obtained from the AEP is generalized to account for coherent attacks by applying the postselection technique \cite{renner2009postselection}. We consider the actual advantage of using different numbers of MUBs by investigating the contribution of finite-size effects. We find that, at fixed dimension, using a higher number of measurement bases yields increased key rates only for high number of rounds, whereas, for a small number, this benefit only preserves for three measurement bases, as the statistical effect of having to estimate a larger number of error rates outweighs the increased key rates for four and more bases.

This paper is structured as follows. In Section~\ref{frame}, we give the basic preliminaries to approach the security proof of a BBM92 scheme. In particular, the state distributed by Eve to Alice and Bob and the error rates are introduced. In Section~\ref{akr}, we optimize the asymptotic key rates for generic dimension and different numbers of MUBs. Then, in Section~\ref{fkr}, we address collective and coherent attacks and find upper bounds on the achievable rates. Finally, in Section~\ref{conc}, we draw our conclusions.

\section{Framework} \label{frame}
Following the general structure of a security proof in QKD \cite{bradler2016finite}, we assume that Eve distributes parts of a global pure quantum state to Alice and Bob, yielding the reduced state $\rho_{AB}$.
Then, we assume that Eve applies a symmetrization map that takes the state $\rho_{AB}$ shared by Alice and Bob into a Bell-diagonal state $\tilde{\rho}_{AB}$.
The reason for this standard assumption is two-fold: First, it can be shown that it is not disadvantageous for Eve to apply it prior to distributing the state, as it only increases her knowledge about the raw key of Alice. 
This implication is well-known in the two-dimensional case. We show in Appendix~\ref{appA} that this is the case also in HD. Second, it leaves the error rates as well as the correlations between Alice and Bob unaffected.

In our HD case, a Bell basis can be defined as follows \cite{wang2017generation}. First, take the $d$-dimensional Bell state
\begin{equation} \label{bellstate}
    \ket{\phi^+} = \frac{1}{\sqrt{d}}\sum_{j=0}^{d-1} \ket{jj}
\end{equation}
and extend it to a basis via 
\begin{equation} \label{bellbasis}
    \ket{\phi_{\alpha,\beta}} = \one\otimes X^\alpha Z^\beta \ket{\phi^+},
\end{equation}
with $\alpha, \beta \in \{0,\ldots,d-1\}$. Note that here we used the Heisenberg-Weyl operators $X^{\alpha}Z^{\beta}$, which generalize the Pauli operators to the HD case. In particular, $X \equiv \sum \ketbraa{j}{j-1}$ is the shift operator and $Z=\sum \omega^j \ketbra{j}$ the clock operator, with $\omega = \exp{(2\pi i/d)}$. We remark that for prime dimensions $d$ the eigenbases of the $d+1$ operators $Z$ and $XZ^k$ with $k\in\{0,1,\ldots d-1\}$ are known to form a maximal set of $d+1$ MUBs \cite{wootters1988,englert2001,durt2010}. For all the other cases, only the bases of $Z$, $X$ and $XZ$ are guaranteed to be mutually unbiased. For prime power dimensions, a similar, modified construction can be used to recover a full set of $d+1$ MUBs \cite{wootters1988,durt2010}.

The symmetrization map applied to $\rho_{AB}$ outputs the state
\begin{align}
    \tilde{\rho}_{AB} &= \frac1{d^2}\sum_{\alpha,\beta=0}^{d-1} [\Lambda_{\alpha,\beta} \otimes \Lambda_{\alpha,-\beta}] \rho_{AB} [\Lambda_{\alpha,\beta}^\dagger \otimes \Lambda_{\alpha,-\beta}^\dagger] \nonumber \\
    &=\frac1{d^2}\sum_{k,l=0}^{d-1}r_{k,l} X^k Z^l \otimes X^k Z^{-l} \label{eq:bdsym} \\
    &= \sum_{\alpha,\beta=0}^{d-1} \lambda_{\alpha,\beta} \ketbra{\phi_{\alpha,\beta}} \equiv \rho_{\text{diag}} \label{diag}
\end{align}
with $\Lambda_{\alpha,\beta}\equiv X^\alpha Z^\beta$. The second line, Eq.~(\ref{eq:bdsym}), is obtained by applying the properties of the Heisenberg-Weyl operators to the decomposition of $\rho_{AB}$ in the Weyl basis, namely
\begin{equation}
    \rho_{AB} = \sum_{k,l,m,n=0}^{d-1}R_{k,l,m,n} X^k Z^l \otimes X^m Z^n,
\end{equation}
and the complex coefficients $R_{k,l,m,n}$ simplify to  $r_{k,l} = R_{k,l,k,-l}$. The indices are assumed to be defined modulo $d$, as $X^d = Z^d = \one$. The last line, Eq.~(\ref{diag}), is proved in Appendix~\ref{appB} and identifies the symmetrized state $\tilde{\rho}_{AB}$ as a Bell-diagonal state $\rho_{\text{diag}}$.

As mentioned above, the symmetrization map leaves the error rates invariant, i.e. $Q_Z(\tilde{\rho}_{AB}) = Q_Z(\rho_{AB})$ and $Q_{XZ^k}(\tilde{\rho}_{AB}) = Q_{XZ^k}(\rho_{AB})$, where
\begin{align}
    Q_Z &\equiv 1-\sum_{j=0}^{d-1} \langle jj|\rho_{AB}|jj\rangle = 1-\sum_{\alpha=0}^{d-1} \lambda_{0,\alpha}, \label{qz} \\
    Q_{XZ^k} &\equiv 1-\sum_{j=0}^{d-1}\langle (\psi^*)_j^k \psi_j^k|\rho_{AB}|(\psi^*)_j^k \psi_j^k\rangle = 1-  \sum_{\alpha=0}^{d-1} \lambda_{\alpha,k\alpha} \label{qxzk}
\end{align}
are the error rates for the eigenbases of $Z$ and $XZ^k$, respectively, and the terms $\lambda_{\alpha,\beta}$ are the coefficients defining $\tilde{\rho}_{AB}$ as a Bell-diagonal state in Eq.~(\ref{diag}), with the indices taken again modulo $d$. In Eq.~(\ref{qxzk}), the eigenbasis $\{|\psi_j^k\rangle\}_{j=1}^d$of $XZ^k$ is given by

\begin{align}\label{eq:eigenvectors}
    \ket{\psi_j^k} &= \begin{cases} \sum_{l=0}^{d-1} \omega^{lj+k\frac{l(l-1)}{2}}\ket{l} & d\text{ odd},\\
        \sum_{l=0}^{d-1} \omega^{lj+k\frac{l^2}{2}}\ket{l} & d\text{ even} \end{cases}
\end{align}
and $\psi^*$ denotes the complex conjugate of $\psi$.

\section{Asymptotic key rates} \label{akr}
We start with the derivation of the achievable asymptotic key rates. As outlined in the previous Section, Eve holds a purification  $\ket{\psi_{ABE}}$ such that the reduced state of Alice and Bob $\trace_E(\ketbra{\psi_{ABE}}) = \tilde{\rho}_{AB}$ is the symmetrized Bell-diagonal state in Eq.~(\ref{diag}). After Alice and Bob performed measurements on their systems, they are left with the classical raw keys $R_A$ and $R_B$, respectively, and the global state is the classical-classical-quantum state $\sigma:=\sigma_{R_AR_BE}$. We evaluate the asymptotic key rate, given by the Devetak-Winter rate \cite{devetak2005distillation}, namely
\begin{equation}\label{eq:devetaksec}
    r_\infty = I(R_A:R_B) - I(R_A:E),
\end{equation}
where $I(X:Y) = H(X)+H(Y)-H(X,Y)$ denotes the mutual information between systems $X$ and $Y$, while $H(X) = -\sum_x\lambda_x\log_2\lambda_x$ is the von Neumann entropy of a system $X$ with eigenvalues $\lambda_x$ and the entropy is evaluated on the corresponding reduced state of $\sigma$.

Rewriting Eq.~(\ref{eq:devetaksec}) in terms of von Neumann entropies, we obtain
\begin{align}\label{eq:devetak2sec}
    I&(R_A:R_B) - I(R_A:E) = H(R_B)-H(E) \nonumber \\
    &-H(R_A,R_B) + H(R_A,E).
\end{align}
For a classical-quantum-quantum state obtained from a global pure state, it holds $H(R_A,E) = H(R_A,B)$, from which
\begin{equation}
    H(R_A,E) \leq H(R_A, R_B)
\end{equation}
follows via the data processing inequality. Thus, Eq.~(\ref{eq:devetak2sec}) can be recast in the following inequality
\begin{align}\label{eq:IIHH}
    I(R_A:R_B) - I(R_A:E) \leq H(R_B) - H(E). 
\end{align}

The right-hand side can be expressed solely in terms of the symmetrized state $\tilde{\rho}_{AB}$: First, note that $H(E)$ is evaluated w.r.t.~the classical-classical-quantum state $\sigma = \sigma_{R_AR_BE}$ which is obtained from the pure state $\ket{\psi}_{ABE}$ by applying the measurements on Alice's and Bob's system. Thus, $H(E)_\sigma = H(E)_{\ket{\psi}_{ABE}} = H(A,B)_{\ket{\psi}_{ABE}} = H(A,B)_{\tilde{\rho}_{AB}}$, as $\tilde{\rho}_{AB} = \trace_E(\ketbra{\psi}_{ABE})$. Second, Bob obtains his raw key bits from a projective rank-1 measurement on his system $\tilde{\rho}_B = \trace_A(\tilde{\rho}_{AB}) = \one / d$, which is therefore completely random and achieves $H(R_B) = \log_2(d) = H(B)_{\tilde{\rho}_{AB}}$.
In order to account for the maximal information Eve can obtain by sending a malicious state $\tilde{\rho}_{AB}$, we have to minimize the right-hand side of Eq.~\eqref{eq:IIHH} over all malicious states $\tilde{\rho}_{AB}$ compatible with observed data:
\opti{min}{\tilde{\rho}_{AB}}{
\log_2(d)- H(A,B)_{\tilde{\rho}_{AB}} \label{optsec}}{Q_i \text{ as observed, }i\in\{Z,X,XZ,\ldots\}.\nonumber}
The optimization in Eq.~(\ref{optsec}) will be performed over the coefficients $\lambda_{\alpha,\beta}$, since the state $\tilde{\rho}_{AB}$ in Eq.~(\ref{diag}) and the error rates in Eqs.~(\ref{qz}) and~(\ref{qxzk}) are expressed in terms of these parameters.

Note that the entropy of $\tilde{\rho}_{AB}$ is determined by the coefficients $\lambda_{\alpha,\beta}$, as the Bell basis is orthogonal.

\subsection{Analytic asymptotic key rates with different numbers of MUBs}
Let us denote by $m$ the number of MUBs that  Alice and Bob measure, i.e., they evaluate the error rates $Q_Z$, $Q_X$, $Q_{XZ}$, ..., $Q_{XZ^{m-2}}$. Depending on $m$, some of the coefficients $\lambda_{\alpha,\beta}$ are fixed while others remain undetermined and have to be optimized over in order to account for the knowledge that Eve might gain. We solve the optimization problem in Eq.~(\ref{optsec}) by constructing the corresponding Lagrangian functional with the constraints given by the error rates $Q_Z$ and $Q_{XZ^k}$, defined in Eqs.~(\ref{qz}) and~(\ref{qxzk}) in terms of the coefficients $\lambda_{\alpha,\beta}$. The set of Equations provided by the constraints can be analytically solved for $m=2$ and $m=d+1$. In Appendix~\ref{appC}, we show that, for $m$ MUBs, the optimal key rate is given by

\begin{align}\label{eq:asymp_lambda}
    r_\infty^{(m)} &= \log_2\frac{d}{d-1} - (m-1)(1-q)\log_2(\eta(d-1)) \nonumber \\
    &\phantom{=}+ (1-Q_Z)\log_2(q-Q_Z-v\eta) \nonumber\\
    &\phantom{=}+ \sum_{k=0}^{m-2}(1-Q_{XZ^k})\log_2(q-Q_{XZ^k}-v\eta)
\end{align}
where 
\begin{align}
    q &:= \left(Q_Z + \sum_{k=0}^{m-2}Q_{XZ^k}\right)/(m-1), \\
    v &:= (d-1)\frac{d-(m-1)}{m-1}
\end{align} and $\eta$ is given by the real solution of the polynomial equation of degree $m$
\begin{align}
    &(d-1)^m (1-q+v\eta) \eta^{m-1} =\nonumber \\
    &(q-Q_Z-v\eta)(q-Q_X-v\eta)\ldots(q-Q_{XZ^{m-2}}-v\eta),
\end{align}
that minimizes the key rate. Let us stress again that this solution is valid for all $d$ if $m\leq 3$. For larger $m$, this is true only for prime dimensions. However, if $m=d+1$, our result holds also in the case of prime power dimensions, as in that case $d+1$ MUBs are known to exist and their measurement would yield complete knowledge about the distributed state.

As mentioned above, while in general there is no closed form solution for $\eta$, there are two special cases which allow for further simplification. 

First, if $m=2$, then $\eta = Q_X Q_Z / (d-1)^2$ and $v = (d-1)^2$, yielding
\begin{align}\label{eq:aympt_asymm_2}
    r_\infty^{(m=2)} =& \log_2d - h(Q_X) - h(Q_Z) \nonumber \\ &- (Q_X+Q_Z)\log_2(d-1)
\end{align}
with the binary Shannon entropy $h(Q) = -Q\log_2Q - (1-Q)\log_2(1-Q)$.
In the symmetric case $Q_X=Q_Z$, the key rate reduces to the asymptotic key rate for two MUBs found in Ref.~\cite{bradler2016finite}.

Second, if $m=d+1$, then $v=0$ and 
\begin{align}\label{eq:aympt_asymm_dp1}
    r_\infty^{(m=d+1)} &=\log_2d + (1-q)\log_2(1-q)  - q\log_2(d-1)  \nonumber \\
    &\phantom{=} + (q-Q_Z)\log(q-Q_Z) \nonumber\\
    &\phantom{=} + \sum_{k=0}^{d-1}(q-Q_{XZ^k})\log_2(q-Q_{XZ^k}).
\end{align}
Note that positivity of the underlying states requires that all error rates are bounded by $q$, implying that the logarithms are evaluated over positive numbers.
If all of the error rates are the same, i.e., $Q_Z=Q_{X}=...=Q_{XZ^{d-1}} \equiv Q$, then $q=(d+1)Q/d$ and the key rate simplifies to
\begin{align}\label{eq:aympt_symm_dp1}
    r_\infty^{(m=d+1)} = \log_2d -h(q) - q\log_2(d^2-1).
    \end{align}

\subsection{Effects of dimension and number of MUBs on asymptotic key and error rates}
\begin{figure}[t]
    \centering
    \includegraphics[width=1\columnwidth]{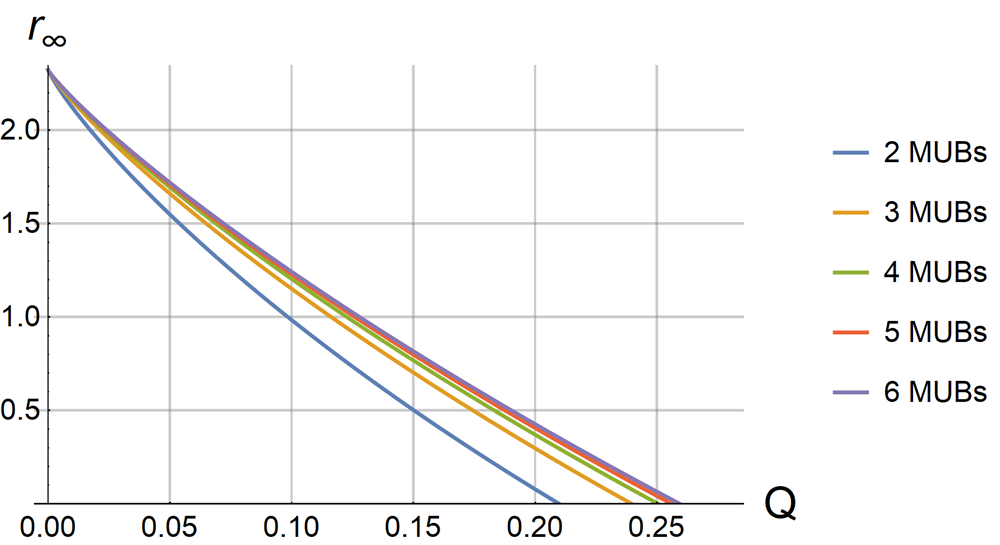}
    \caption{Asymptotic key rates for five-dimensional systems and all possible numbers of measured MUBs, under the assumption that the error rates in all bases are the same. For $m=2$ and $m=d+1$, closed formulas are given by Eqs.~(\ref{eq:aympt_asymm_2}) and (\ref{eq:aympt_symm_dp1}), respectively.} 
    \label{fig:symmd5}
\end{figure}

\begin{figure}[t]
    \centering
    \includegraphics[width=1\columnwidth]{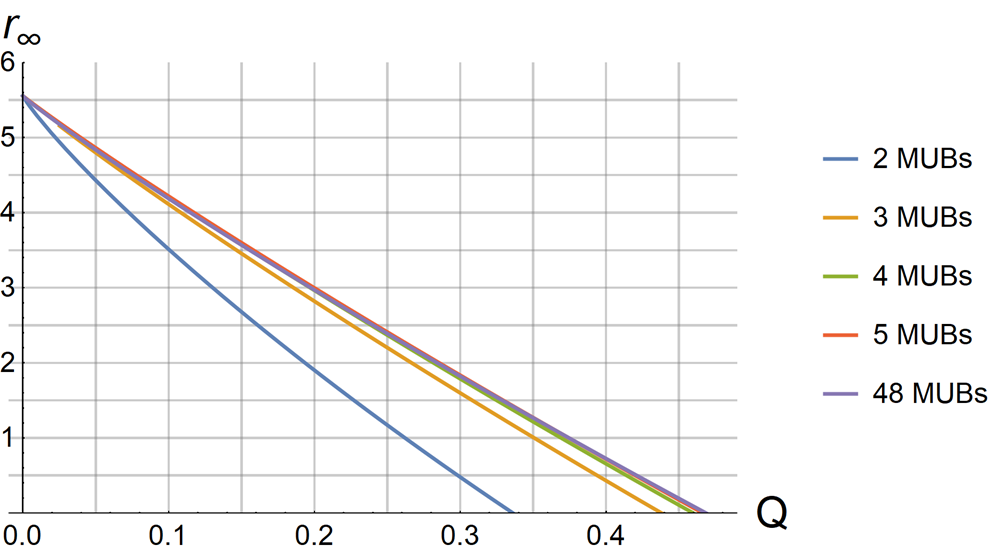}
    \caption{Asymptotic key rates for systems with $d=47$ and different numbers of measured MUBs, under the assumption that the error rates in all bases are the same. For $m=2$ and $m=d+1$, closed formulas are given by Eqs.~(\ref{eq:aympt_asymm_2}) and (\ref{eq:aympt_symm_dp1}), respectively.} 
    \label{d47}
\end{figure}

In Fig.~\ref{fig:symmd5} we display the asymptotic key rates for $d=5$ and $m=2,\ldots,d+1$ in the case of symmetric error rates, i.e. when $Q_Z = Q_{XZ^{k}} \,\, \forall\,k=0,1,\ldots,d-1$. As expected, we see that, for fixed dimension, the maximum tolerable error rate is improved by enhancing the number of mutually unbiased measurements. However, here we also note that the enhancement of the maximum tolerable error rate monotonically decreases as the number of MUBs increases. This observation suggests that one does not really need to exploit all of the $d+1$ MUBs allowed by the dimension of the system and that one already gets a good advantage with three MUBs. To stress this point, we also show in Fig.~\ref{d47} the secret key rate for a large prime dimension, namely $d=47$, and remark that the improvement achieved by increasing the number of MUBs from three to $d+1$ is smaller than the one obtained moving from $m=2$ to $m=3$.  

\begin{figure}[t]
    \centering
    \includegraphics[width=1\columnwidth]{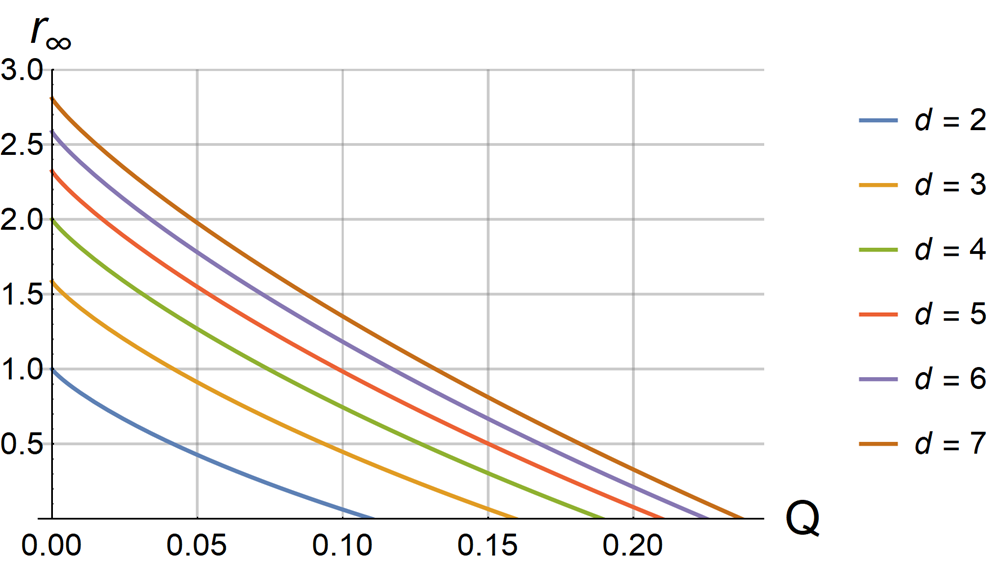}
    \caption{Asymptotic key rates for $d$-dimensional systems and two MUBs under the assumption that the error rates in all bases are the same. The analytical expression is given in Eq.~(\ref{eq:aympt_asymm_2}) after setting $Q_X=Q_Z$.} 
    \label{fig:symmk2}
\end{figure}

\begin{figure}[t]
    \centering
    \includegraphics[width=1\columnwidth]{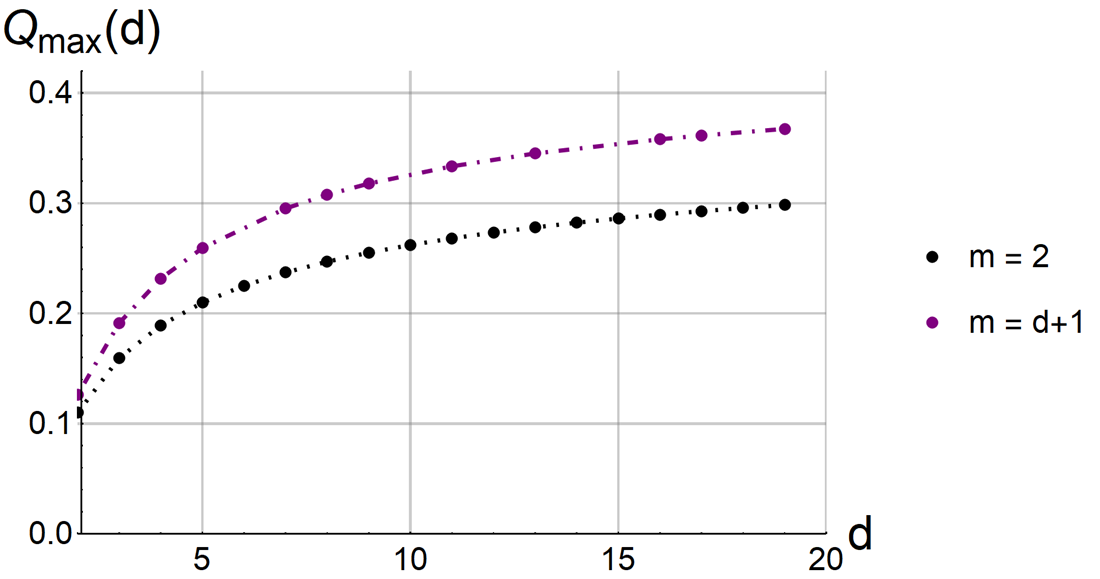}
    \caption{Maximum tolerable error rates $Q_{\text{max}}$ as a function of the dimension $d$ of the systems for the cases $m=2$ (black points connected by black dotted line) and $m=d+1$ (purple points connected by purple dashed-dotted line).} 
    \label{maxQ}
\end{figure}

In Fig.~\ref{fig:symmk2}, we display the key rate for two MUBs and different dimensions with symmetric error rates. Note that, by increasing the dimension, we both get higher key rates and a larger tolerance of noise. This second advantage is highlighted in Fig.~\ref{maxQ}, where we plot, in the symmetric case, the maximum tolerable error rate $Q_{\text{max}}$, namely the error rate at which the key rate becomes zero, as a function of the dimension $d$ of the system. There, we plot the same quantity also for the case $m=d+1$, which shows again the enhancement that one gets by exploiting more than two MUBs, as displayed in Figs.~\ref{fig:symmd5} and~\ref{d47} for the particular cases $d=5$ and $d=47$. 

\begin{figure}[t]
    \centering
    \includegraphics[width=1\columnwidth]{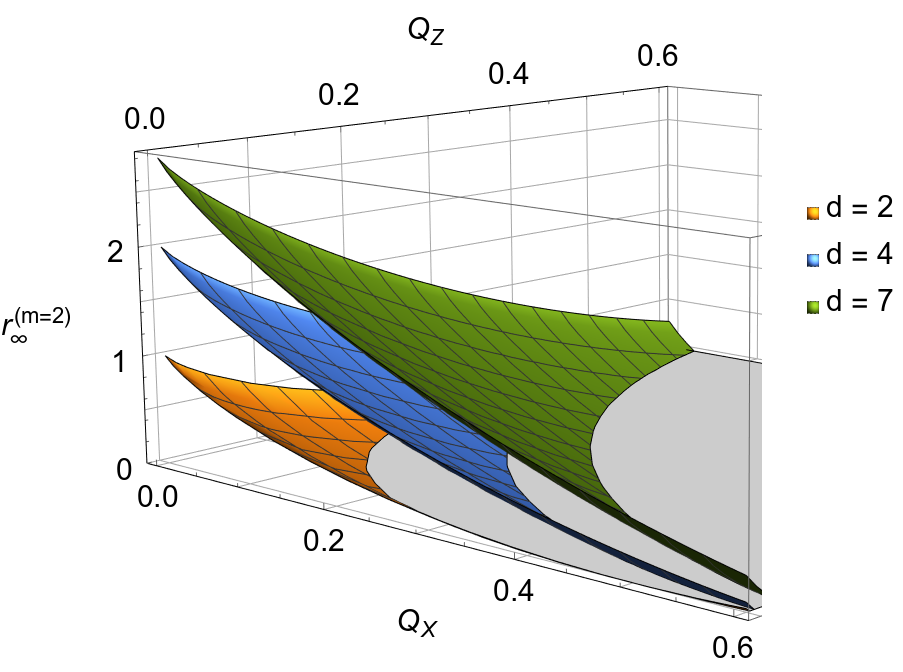}
    \caption{Asymptotic key rates for $d$-dimensional systems and two MUBs as a function of the two error rates $Q_X$ and $Q_Z$. The analytical expression is given in Eq.~(\ref{eq:aympt_asymm_2}).} 
    \label{fig:asymk2_1}
\end{figure}

Finally, we address the case of asymmetric errors in Fig.~\ref{fig:asymk2_1}, where we display the key rates for three different choices of $d$ and two MUBs.
Note, however, that in real world implementations one would expect that implementation of a higher number of MUB measurements or higher dimensional schemes leads to higher error rates. This ultimately leads to a trade off between the increased rates and tolerable error rates when using higher dimensional encodings and multiple measurement bases, and the accompanying experimental challenges.


\section{Finite key rates} \label{fkr}
We now turn to the case of finite key rates \cite{renner2008security}, namely the number of rounds $N$ of the protocol is finite. We recall that Eve is assumed to hold the purifying system $|\psi_{\text{ABE}}\rangle$. In this scenario, we can address the most general attacks that Eve can perform, which are known as \textit{collective} and \textit{coherent} attacks \cite{biham1997,renner2005,scarani2009,grasselli2020quantum}. The main difference between the two lies in the structure of the global state $\tilde{\rho}_{\text{AB}}^{(N)}$ distributed to Alice and Bob, accounting for all the $N$ rounds. In the case of collective attacks, this is a tensor product of every single-round state (i.i.d.), i.e. $\tilde{\rho}_{\text{AB}}^{(N)}=\tilde{\rho}_{\text{AB}}^{\otimes N}$. On the contrary, for coherent attacks in general $\tilde{\rho}_{\text{AB}}^{(N)}\neq\tilde{\rho}_{\text{AB}}^{\otimes N}$, allowing for correlations between single-round states. 
For permutation invariant protocols the key rates that are secure against collective attacks asymptotically coincide with key rates that are secure against coherent attacks \cite{scarani2009,sheridan2010rapid,grasselli2020quantum}.
In the latter case the state is in general a convex combination of i.i.d. states \cite{renner2007symmetry}, which one needs to approximate by a single tensor-product state in order to bound the key rates through the asymptotic equipartition properties expressed in terms of von Neumann entropies \cite{renner2007symmetry,renner2008security,sheridan2010njp,bradler2016finite}, as for the collective attacks. This approximation can be achieved by using the so-called postselection technique \cite{renner2009postselection}. Alternatively, one can exploit suitable entropic uncertainty relations (EURs), but their generalization to more than two MUBs is still a subject of research.

Here we consider both, collective and coherent attacks. First, we address the security against collective attacks. We compare bounds on the secret key rates derived from the EURs \cite{tomamichel2011uncertainty} in the case $m=2$ and from the asymptotic equipartition property \cite{renner2008security} for $m\geq 2$. Last, we consider the case of coherent attacks and exploit the results obtained in Ref.~\cite{sheridan2010njp} from the postselection technique \cite{renner2009postselection}.

The steps of the protocol are the standard ones of an entanglement-based BB84-like scheme, namely BBM92 like \cite{bennett1992}, i.e. distribution, measurement, sifting, parameter estimation, error correction and privacy amplification \cite{mink2012,tomamichel2012tight,tomamichel2017,grasselli2020quantum}. 

\subsection{Upper bounds on finite key rates}

We begin with the case of two MUBs, that is, Alice and Bob measure either in the $Z$ ($n$ times) or in the $X$ basis ($k$ times), yielding a total of $N=n+k$ measurement steps. Without loss of generality, we assume that they choose the $Z$ basis for the key generation and the $X$ basis as a test basis. The $X$-basis measurements are then used to compute $Q_X$, see Eq.~(\ref{qxzk}). Before they start the protocol, Alice and Bob set a maximum error tolerance $Q_{\text{tol}}$ such that, if the parameter estimation outputs $Q_X>Q_{\text{tol}}$, they abort the protocol \cite{tomamichel2012tight}. We will refer to the security parameters related to error correction and to privacy amplification as $\epsilon_{\text{EC}}$ and $\epsilon_{\text{PA}}$, respectively.

Then, during error correction, Alice and Bob have to reveal a certain number $\leakEC$ of bits. Consequently, Alice computes a hash bitstring of length given by $\lceil\log_2(1/\epsilon_{\text{EC}})\rceil \leq \log_2(2/\epsilon_{\text{EC}})$ bits from her raw key and sends the hash to Bob, who compares it with the hash of his own key in order to make sure that the error correction procedure worked. Finally, in the privacy amplification step, Alice and Bob apply another hash function to reduce the length of the key to \cite{tomamichel2012tight,grasselli2020quantum}
\begin{equation}
\label{eq:extraction}
    l \leq H_\text{min}^\epsilon(Z_A^n|E) - \leakEC - \log_2\frac{2}{\epsilon_\text{EC}}-2\log_2\frac1{2\epsilon_\text{PA}}
\end{equation}
from which the finite key rate $r = l/n$ can be deduced. In Eq.~(\ref{eq:extraction}), $Z^n$ is the key bitstring of length $n$, $H_\text{min}^\epsilon(Z^n|E)$ denotes the conditional smooth min-entropy \cite{renner2005smoothent,tomamichel2010,tomamichelbook} with smoothing parameter $\epsilon$, and $\epsilon_{\text PA}$ is a security parameter related to the length of the hash function used in the privacy amplification step \cite{grasselli2020quantum}. It can be shown \cite{grasselli2020quantum} that the extracted key in Eq.~(\ref{eq:extraction}) is $\epsilon_\text{tot}$-secure with $\epsilon_\text{tot} = \epsilon + \epsilon_\text{PA} + \epsilon_\text{EC}$.

In order to bound $H_\text{min}^\epsilon(Z^n|E)$, we use the following uncertainty relation for smooth entropies \cite{tomamichel2012tight,tomamichel2011uncertainty}
\begin{equation}\label{eq:uncertainty}
    H_\text{min}^\epsilon(Z_A^n|E) \geq n C - H_\text{max}^\epsilon(X_A^k|X_B^k). 
\end{equation}
Here, $C$ denotes the incompatibility of the measurements and is given by $C:= -\log_2c$ where $c:=\max_{i,j} \Vert \sqrt{M_i}\sqrt{N_j}\Vert_\infty^2$ with $M_i$ and $N_j$ denoting the POVM elements of the two different measurement bases. Thus, in case of perfect projective measurements in the $Z$ and the $X$-basis, the incompatibility is given by $c = \frac1d$ and therefore $C=\log_2d$. However, in a real world implementation, the measurements will not be perfect, which is why we leave $C$ as a free parameter that has to be determined for the specific setup in use.

The smooth max-entropy $H_\text{max}^\epsilon(X_A^k|X_B^k)$ identifies the amount of information that one needs to find the value of the string $X_A^k$ from a given string $X_B^k$. 
 This is related to the maximum error tolerance $Q_{\text{tol}}$ and includes statistical uncertainties. 
Following Ref.~\cite{tomamichel2012tight} and the related Supplemental Material, by exploiting Serfling's bound for the sum in sampling without replacement \cite{serfling1974}, this can be accounted for by increasing the maximum error tolerance $Q_{\text{tol}}$ by 
\begin{equation}\label{eq:mueps}
    \mu_{\epsilon'} = \sqrt{\frac{N(\tilde{k}+1)\ln(1/\epsilon')}{n\tilde{k}^2}}. 
\end{equation}
Here, $\tilde{k} = k/m$ accounts for the fact that we have to split th $k$ parameter estimation rounds into $m$ blocks in order to estimate the $m$ different error rates. The parameter $\epsilon'$ is proportional to the smoothing parameter $\epsilon$. In particular, $\epsilon' = \epsilon\sqrt{p_{\text{pass}}}$, where $p_{\text{pass}}$ identifies the probability that the correlation test between $X_A^k$ and $X_B^k$ passes. For the sake of simplicity, in the following we assume $p_{\text{pass}}=1$, and then $\epsilon'=\epsilon$.
Hence, the smooth max-entropy can be shown 
\cite{tomamichel2012tight} 
to be upper bounded by 
\begin{align} \label{smoothmax}
    H_\text{max}^\epsilon(X_A^k|X_B^k) &\leq \log_2 \sum_{l=0}^{\lfloor n(Q_{\text{tol}}+\mu_{\epsilon})\rfloor} \binom{n}{l}(d-1)^l \nonumber \\
    &\leq n[h(Q_{\text{tol}}+\mu_{\epsilon}) + (Q_{\text{tol}}+\mu_{\epsilon})\log_2(d-1)] 
\end{align}
for $Q_{\text{tol}}+\mu_{\epsilon} \leq \frac12$. By inserting the uncertainty relation and the bound on the max-entropy into Eq.~(\ref{eq:extraction}), we get
\begin{align}
    r&(\epsilon, \epsilon_\text{EC}, \epsilon_\text{PA},n,k) \leq C-h(Q_{\text{tol}}+\mu_{\epsilon})-(Q_{\text{tol}}+\mu_{\epsilon})\nonumber \\
     & \cdot\log_2(d-1)-\frac1n\left(\leakEC +\log_2\frac{2}{\epsilon_\text{EC}}+2\log_2\frac1{2\epsilon_\text{PA}}\right).
\end{align}
The total rate is then given by $\frac nN r$ with $n=N-k$ and can be optimized over $k, \epsilon, \epsilon_\text{EC}$ and $\epsilon_\text{PA}$:
\begin{equation}
    \hat{r}(\epsilon_\text{tot},N) = \max_{k,\epsilon,\epsilon_\text{EC},\epsilon_\text{PA}} \hat{r}(\epsilon_\text{tot},N,k),
\end{equation}
with 
$\epsilon_\text{tot} \geq 2\epsilon + 2\sqrt{2\epsilon_\text{EC}} + \epsilon_\text{PA}$ \cite{nahar2024postselection}. 

Finally, note that the optimal value of $\leakEC$ is given by $\leakEC = h(Q) + Q\log_2(d - 1)$. For a finite number of rounds, however, this asymptotic value is not achievable and will be slightly worse, depending on the implemented error correction scheme. In practice, one accounts for this by multiplying the asymptotic value by some factor $f$ (e.g. $f=1.1$).

In the case of more than two MUBs, the uncertainty relation in Eq.~(\ref{eq:uncertainty}) is not directly applicable, and the existing relations in terms of multiple MUBs \cite{wang2021} have some issues that still have to be resolved. Instead, we use the asymptotic equipartition property (AEP), reading \cite{tomamichel2009}
\begin{equation} \label{aep}
    \frac1nH_\text{min}^{\epsilon}(Z_A^n|E) \geq H(Z_A|E)_{\tilde{\rho}_{AB}} - \frac{4}{\sqrt{n}}\log_2 (2 + \sqrt{d})\sqrt{\log_2\frac{2}{\epsilon^2}}.
\end{equation}
We now assume symmetric errors, i.e., $Q\equiv Q_Z = Q_{XZ^k} \,\,\forall\,k=0,\ldots,d-1$. Then,
plugging the AEP into Eq.~(\ref{eq:extraction}) we get the achievable key rate as follows 
\begin{align}
    \hat{r}&(\epsilon_\text{tot}, N, n, m, Q) =\frac nN[\underset{r_\infty^{(m)}(Q_{\text{tol}}+\mu_{\epsilon})}{\underbrace{\min_{\tilde{\rho}_{AB}}(H(Z_A|E) - \operatorname{leak}_\text{EC})}}]\nonumber\\
    &-\frac{1}{N}\left[\log_2\frac{1}{2\epsilon_\text{EC}\epsilon_\text{PA}^2} + 4\sqrt{n}\log_2 (2 + \sqrt{d})\sqrt{\log_2\frac{2}{\epsilon^2}}\right],
    \label{AEPbound}
\end{align}
which still can be maximized over the test rounds $k$ and the security parameters, satisfying $\epsilon_\text{tot} \geq 2\epsilon + 2\sqrt{2\epsilon_\text{EC}} + \epsilon_\text{PA}$ \cite{nahar2024postselection}. 

Note that we can apply again the argument provided by the Serfling bound \cite{serfling1974} that we used to define the statistical correction $\mu_{\epsilon}$ in Eq.~(\ref{eq:mueps}) and that we already employed to upper bound the smooth max-entropy in Eq.~(\ref{smoothmax}). 
By doing so, we can replace the optimization over the single run joint state in Eq.~(\ref{AEPbound}) with the asymptotic key rate obtained in the previous Section, here to be expressed as a function of $Q_{\text{tol}}+\mu_{\epsilon}$.

Only for $m=2$ and $m=d+1$ (if $d$ is a power of a prime) MUBs we can obtain the closed form solutions from Eqs.~(\ref{eq:aympt_asymm_2}) with $Q_X=Q_Z$, and~(\ref{eq:aympt_symm_dp1}), respectively. In the other cases, we need to find the roots of the polynomial in Eq.~(\ref{eq:conditionlambda}), but a numerical evaluation is straightforwardly achieved.

Finally, we address the case of coherent attacks. To this aim, we exploit a revised version of the postselection technique \cite{renner2009postselection}, following the analysis detailed in \cite{nahar2024postselection}.

There, the authors show that the security of a protocol which is $\epsilon_\text{tot}$-secure against collective attacks can be improved to get a key rate $\epsilon_\text{coh}$-secure against coherent attacks by setting
\begin{align}
    \epsilon_\text{coh} = \epsilon_\text{tot}\binom{N+d^4-1}{N},
\end{align}
thus damping the resulting key rate as follows
\begin{align} \label{postselection}
    r_\text{coh}(\epsilon_\text{coh}) = r_\text{col}(\epsilon_\text{tot}) - \frac{2}{N}\log_2\binom{N+d^4-1}{N}.
\end{align}

\subsection{Comparison of finite key rates for different numbers of MUBs}
\begin{figure*}
    \centering
    \includegraphics[width=2\columnwidth]{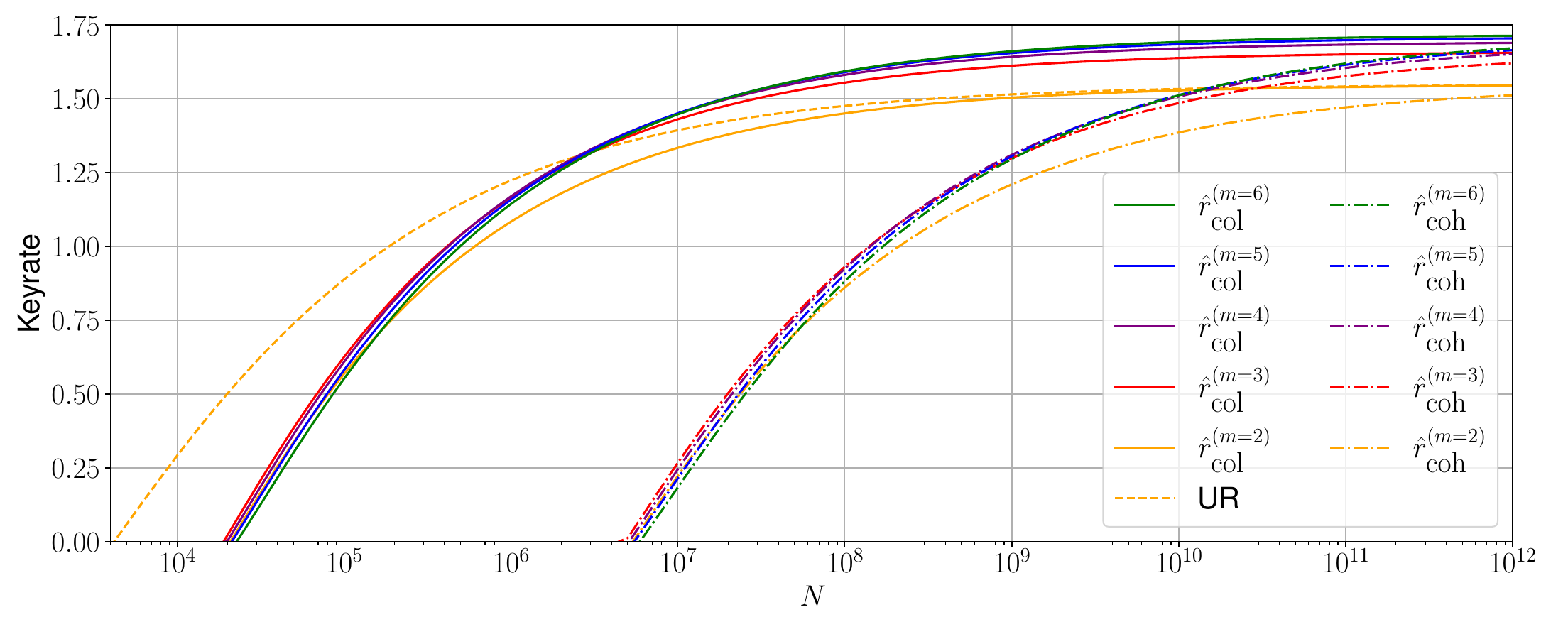}
    \includegraphics[width=0.99\columnwidth]{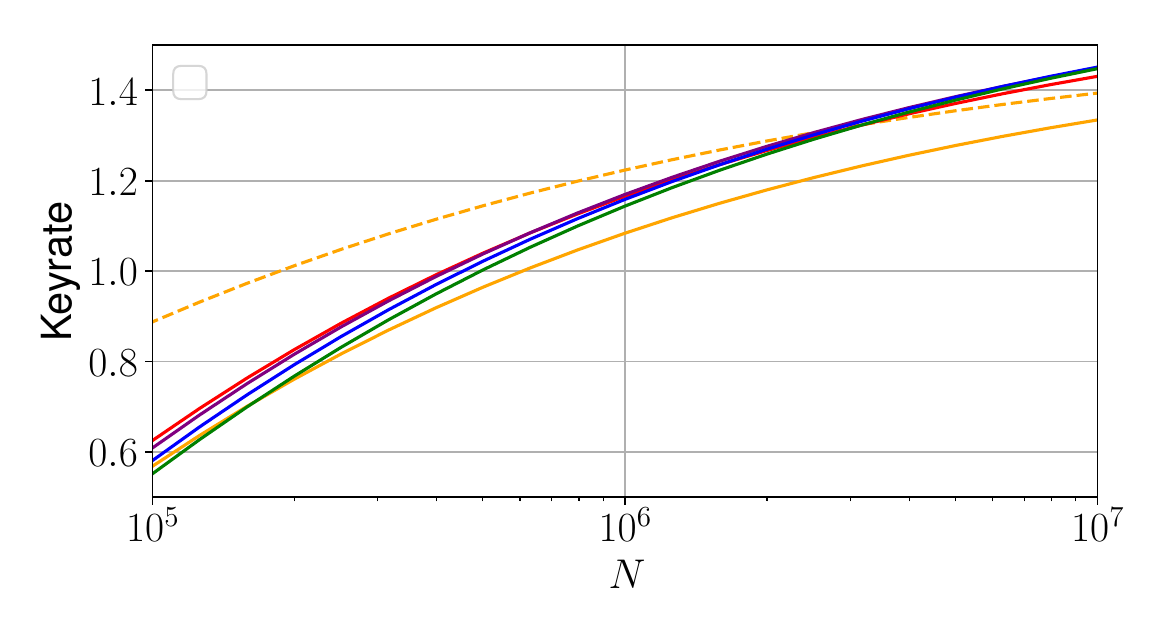}
    \includegraphics[width=0.99\columnwidth]{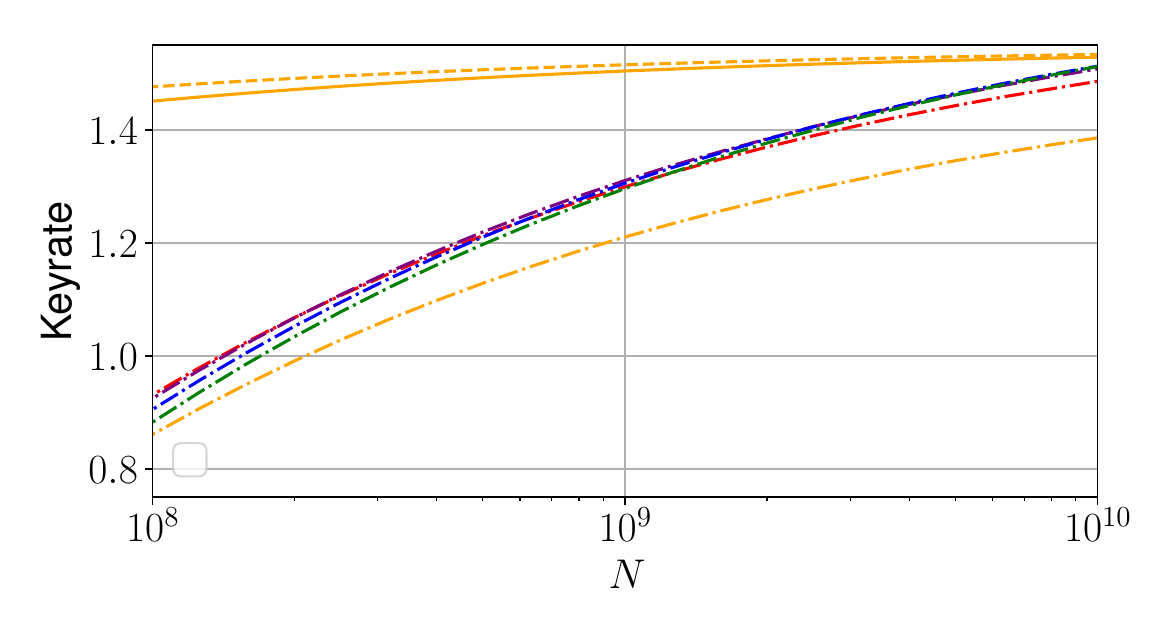}
   \caption{Top: Finite key rates secure against collective (solid lines, Eq.~(\ref{aep})) and coherent (dot-dashed lines, Eq.~(\ref{postselection})) attacks for $d=5$,  $Q=0.05$, $\epsilon_\text{tot} = 10^{-10}$ and different numbers of MUBs $m$, each optimized over the number of parameter estimation rounds $k$ and the tolerated errors $\epsilon$, $\epsilon_\text{C}$ and $\epsilon_\text{PA}$. The dashed yellow line indicates the achievable key rate using standard two-bases BB84 with the EUR in Eq.~(\ref{eq:uncertainty}). Bottom: zoom onto the regions where using more than three MUBs starts to yield an advantage assuming collective (left) and coherent (right) attacks. } 
    \label{fig:finitekey}
\end{figure*}

In order to assess the achievable secure key rates under the assumption of collective attacks, we maximize the rate in Eq.~(\ref{AEPbound}) for fixed $d=5$, $Q=0.05$ and $\epsilon_\text{tot} = 10^{-10}$ for different $N$ and $m$ over $k$, $\epsilon$, $\epsilon_\text{EC}$ and $\epsilon_\text{PA}$. For coherent attacks, we employ the postselection technique in Eq.~(\ref{postselection}) instead. Note that we use the asymptotic value for the error correction term  $\leakEC(Q+\mu_\epsilon)$ by setting the factor $f=1$. While this choice probably underestimates the number of bits required for error correction slightly, we stress that this is probably compensated for by using the asymptotic value for the worst error rate given by $Q+\mu_\epsilon$ instead of the more probable $Q$. In proper implementations of the protocol, the leakage would be replaced by an actual count of transmitted bits instead.
The result of the optimization is displayed in Fig.~\ref{fig:finitekey} for both kinds of attacks and different numbers of MUBs. 

In order to compare the obtained rates with state-of-the-art rates for two measurement bases using EURs, we display the standard BB84 rate found from EURs as a dashed, yellow line, while the solid lines show the key rates obtained from the AEP with $m = 2,\ldots ,d+1$. Note that for fixed $m=2$, the key rate derived from the EUR is globally better than the one found with the AEP both in terms of length of the key per number of rounds and in terms of minimum number of signals to get a positive rate. In fact, only for a large number of rounds we find higher achievable key rates using multiple bases. This result hints that the search for EURs with more than two MUBs promises protocols with higher noise tolerance and higher key rates.

If we focus on the key rates derived from the AEP for large $N$, we note that, as found in the asymptotic case, exploiting more than two MUBs allows to achieve higher rates. We remark that in the case $m=3$ our optimization provides a key rate comparable with the one found in Ref.~\cite{bradler2016finite} for the same values of the parameters $d$, $m$, $Q$ and $\epsilon_{\text{tot}}$. There, they use the AEP as well, but they also explore a different estimate based on finite block length quantum coding \cite{tomamichel2016}.

An interesting trade off can be seen when comparing the obtained rates for different numbers of bases: While taking more bases into account leads to a larger asymptotic rate, the fact that the number of parameter estimation rounds has to be split into $m$ blocks to estimate more QBERs reduces the rate in the finite key regime sufficiently to actually obtain smaller rates for small $N$. Unexpectedly, if $N$ is small enough, the optimal choice for the number of MUBs is $m=3$. In particular, only with $N \gtrsim 3\cdot 10^6$ rounds for collective attacks (and $N\gtrsim 10^9$ for coherent attacks) it makes sense to use more than three MUBs.

\section{Conclusions} \label{conc}
We devised a complete analysis of the security of a BBM92-like protocol for generic dimension and for every allowed number of MUBs. Our work provides the proof that in the asymptotic regime the secret key rates and the maximum tolerable error rates grow as the number of exploited number of MUBs increases. Quite surprisingly, in the finite-key scenario we find that, if the number of rounds is not too large, it is optimal to use three MUBs.

In the asymptotic regime we found the analytic expression of the key rate when $d+1$ MUBs are used and provided a numerical optimization for all the other cases. For a fixed dimension, the relative improvement in the maximum tolerable error rate $Q_{\text{max}}$ that one gets by choosing $m+1$ MUBs instead of $m$ is a decreasing monotonic function of $m$. Interestingly, the difference between $Q_{\text{max}}(m=3)$ and $Q_{\text{max}}(m=2)$ is significant. 

In the case of a finite number of resources, we derived achievable upper bounds on the finite key rates against collective and coherent attacks. For $m=2$ and $m=3$ MUBs, our key rates are compatible with the ones recently retrieved in Ref.~\cite{bradler2016finite}. We remark that, in the the case $m=2$ and generic dimension, the known bound obtained from the EUR outperforms the one that we retrieved from the AEP. Together with the fact that the postselection technique probably yields too optimistic key rates \cite{nahar2024postselection}, it is clear that a further improvement of the performance of BBM92-like protocols needs to be obtained in the future by retrieving and exploiting uncertainty relations for smooth conditional entropies with more than two MUBs. 

\acknowledgments

We thank Shlok Nahar for making us aware of the flaws in the original postselection results. GC and CM acknowledge the EU H2020 QuantERA ERA-NET Cofund in Quantum Technologies project QuICHE and support from the PNRR MUR Project PE0000023-NQSTI.

\appendix

\section{Effect of the symmetrization map} \label{appA}

Here we prove that the application of the symmetrization map in Eq.~(\ref{eq:bdsym}) by Eve can only increase her knowledge of the key, i.e., $H(R_A | E)_{\tilde{\rho}_{AB}} \leq H(R_A | E)_{\rho_{AB}}$. The proof is a generalization to $d$-dimensional systems of the one known for the 2D case (see for instance Appendix 3.5 in Ref.~\cite{grasselli2020quantum}).

Alice and Bob actively symmetrize the distributed states by drawing randomly the numbers $0\leq k,l \leq d-1$ and, depending on the outcome, rotate the state according to
\begin{align}
    X^k Z^l \otimes X^k Z^{-l} \rho_{AB} (X^k Z^l \otimes X^k Z^{-l})^\dagger.
\end{align}
As Alice and Bob have to communicate their choice of $k$ and $l$, these numbers are known to Eve and we assume that she holds the corresponding purification $\ket{\phi_{ABE}^{k,l}}$ for each choice of $k$ and $l$.
The choice of $k$ and $l$ is stored in the register $T$, yielding Eve's state
\begin{equation} \label{eq:evesstate}
    \tilde{\rho}_{ABET} = \frac1{d^2}\sum_{k,l}\ketbra{\phi_{ABE}^{k,l}}\otimes \ketbra{k,l}_T.
\end{equation}
This state can be purified as well and yields Eve's global state
\begin{equation}    \ket{\phi_{ABETT^\prime}} = \frac1d\sum_{k,l}\ket{\phi_{ABE}^{k,l}} \otimes \ket{k,l}_T \otimes \ket{k,l}_{T^\prime}.
\end{equation}
Now, we can exploit the strong sub-additivity property for conditional entropies and write
\begin{align} \label{eq:strongsub}
    H(R_A|ETT^\prime)_{|\phi\rangle} \leq H(R_A|ET)_{\tilde{\rho}}.
\end{align}
where $\tilde{\rho}$ is the state shared by Eve and Alice after that Alice applies her measurement map, namely 
\begin{align}
    \tilde{\rho}_{R_AET} = \frac1{d^2}\sum_{k,l}\rho_{R_AE}^{k,l}\otimes \ketbra{k,l}_T.\label{eq:rhotraet}
\end{align}
Note that, being the $T$-part of Eq.~(\ref{eq:rhotraet}) classical, its entropy can be written as
\begin{align}\label{eq:Hassum}
H(R_A|ET)_{\tilde{\rho}} = \frac1{d^2}\sum_{k,l} H(R_A|E)_{\rho_{R_AE}^{k,l}},
\end{align}
with the short-hand notation $\tilde{\rho} = \tilde{\rho}_{R_AET}$.
However, the entropy of $\rho_{R_AE}^{k,l}$ is independent of $k,l$. This can be seen as follows: Fix $k$ and $l$ and consider the spectral decomposition of $\rho_{AB}$,
\begin{equation}
    \rho_{AB} = \sum_{\lambda} \lambda \ketbra{\lambda}.
\end{equation}
Then, $\ket{\phi_{ABE}^{k,l}}$ can be written as 
\begin{equation}
    \ket{\phi_{ABE}^{k,l}} = \sum_\lambda \sqrt{\lambda} \ket{\lambda^{k,l}}_{AB} \otimes \ket{e_\lambda}_E,
\end{equation}
with $\ket{\lambda^{k,l}} = X^kZ^l \otimes X^kZ^{-l} \ket{\lambda}$.
Thus, we can write explicitly:
\begin{align}
    &\rho_{R_AE}^{k,l} = \\
    =&\sum_{a=0}^{d-1} \ketbra{a} \otimes \sum_{\lambda, \mu} \sqrt{\lambda\mu}\trace_B[\langle a|\lambda^{k,l}\rangle\langle \mu^{k,l}|a\rangle]\ketbraa{e_\lambda}{e_\mu} \nonumber \\
    =& \sum_{a=0}^{d-1} \ketbra{a} \otimes \sum_{\lambda, \mu} \sqrt{\lambda\mu}\trace_B[\langle a|(\Lambda_{k,l} \otimes \Lambda_{k,-l})|\lambda\rangle \nonumber \\
    &\langle\mu|(\Lambda_{k,l}\otimes \Lambda_{k,-l})^\dagger|a\rangle]\ketbraa{e_\lambda}{e_\mu} \nonumber \\
    =& \sum_{a=0}^{d-1} \ketbra{a} \otimes \sum_{\lambda, \mu} \sqrt{\lambda\mu}\trace_B[\langle a|(\Lambda_{k,l} \otimes \one)|\lambda\rangle \nonumber \\
    &\langle \mu|(\Lambda_{k,l}\otimes \one)^\dagger|a\rangle]\ketbraa{e_\lambda}{e_\mu} \nonumber \\
    =& \sum_{a=0}^{d-1} \ketbra{a} \otimes \sum_{\lambda, \mu} \sqrt{\lambda\mu}\trace_B[\langle a-k|\lambda\rangle\langle \mu|a-k\rangle]\ketbraa{e_\lambda}{e_\mu} \nonumber \\
    =& \sum_{a=0}^{d-1} \ketbra{a+k} \otimes \sum_{\lambda, \mu} \sqrt{\lambda\mu}\trace_B[\langle a|\lambda\rangle\langle \mu|a\rangle]\ketbraa{e_\lambda}{e_\mu} \nonumber \\
    =:& \sum_{a=0}^{d-1} \ketbra{a+k} \otimes \rho_E^a
\end{align}
with $\Lambda_{k,l}\equiv X^kZ^l$.
Then, we see that the result depends only on $k$, which can be compensated by a relabelling of the classical outcomes of Alice's measurements. This relabelling is ignored by the entropy, leading to
\begin{equation}
    H(R_A|E)_{\rho_{R_AE}^{k,l}} = H(R_A|E)_{\rho_{R_AE}^{0,0}} = H(R_A|E)_{\rho_{AB}} \quad \forall k,l.
\end{equation}
Therefore, from Eq.~(\ref{eq:Hassum}) we obtain 
\begin{align} \label{fin}
H(R_A|ET)_{\tilde{\rho}} = H(R_A|E)_{\rho_{AB}}.
\end{align}
Hence, by putting together Eqs.~(\ref{eq:strongsub}) and~(\ref{fin}) the claim follows.

\section{The symmetrized state is Bell-diagonal}\label{appB}
Here we want to prove the equivalence between Eq.~(\ref{eq:bdsym}) and Eq.~(\ref{diag}), namely that the symmetrization map outputs a Bell-diagonal state.

This is straightforward, since a Bell-diagonal state $\rho_{\text{diag}}$ can be expressed as a convex combination of the projectors onto the basis states in Eq.~(\ref{bellbasis}), i.e.
\begin{align}\label{eq:rhoBD}
    \rho_{\text{diag}} &= \sum_{\alpha,\beta=0}^{d-1} \lambda_{\alpha,\beta} \ketbra{\phi_{\alpha,\beta}} \nonumber \\
 &= \sum_{\alpha,\beta=0}^{d-1} \lambda_{\alpha,\beta} [\one \otimes X^\alpha Z^\beta] \ketbra{\phi^+} [\one \otimes (X^\alpha Z^\beta)^\dagger] \nonumber \\
    &= \frac1{d^2}\sum_{k,l} (\sum_{\alpha,\beta} \lambda_{\alpha,\beta}\omega^{\alpha l - \beta k}) X^k Z^l \otimes X^k Z^{-l} \nonumber \\
    &= \frac1{d^2}\sum_{k,l} r_{k,l} X^k Z^l \otimes X^k Z^{-l},
\end{align}
where the third line is obtained by expanding the projector $\ketbra{\phi^+}$ according to Eq.~(\ref{bellstate}) and applying the Weyl-Heisenberg operators. 
Note that this distributed state is the same obtained in Ref.~\cite{bradler2016finite} through the Choi-Jamiolkowski isomorphism.

\section{Optimization of the asymptotic key rate}\label{appC}

Here we derive the asymptotic key rate for $m$ MUBs reported in Eq.~(\ref{eq:asymp_lambda}).

The Lagrange functional corresponding to the optimization problem in Eq.~(\ref{optsec}) can be written as
\begin{align} \label{lagrangian}
    \mathcal{L} &= \log_2d +\sum_{\alpha,\beta=0}^{d-1} \lambda_{\alpha,\beta} \log_2 \lambda_{\alpha,\beta} - \mu_Z(1-\sum_{\alpha=0}^{d-1} \lambda_{0,\alpha}-Q_Z) \nonumber \\
    &\phantom{=}- \sum_{k=0}^{m-2} \mu_k(1-\sum_{\alpha=0}^{d-1} \lambda_{\alpha,k\alpha} - Q_{XZ^k}) - \mu_\text{N}(1-\sum_{\alpha,\beta=0}^{d-1} \lambda_{\alpha,\beta}).
\end{align}
The additional constraints on the positivity of the coefficients $\lambda_{\alpha,\beta}$ are not stated explicitly, but we will find that the optimal solution will satisfy these constraint inherently.

From the Lagrangian, we obtain the following constraints:
\begin{align}
    \frac{\partial\mathcal{L}}{\partial\lambda_{0,0}} &= \log_2 \lambda_{00} +\frac{1}{\ln2}+\mu_Z+\sum_{k=0}^{m-2}\mu_k+\mu_\text{N}=0,&  \label{eq:lag1}\\
    \frac{\partial\mathcal{L}}{\partial\lambda_{0,i}} &= \log_2 \lambda_{0i} +\frac{1}{\ln2}+\mu_Z+\mu_\text{N}=0 \label{eq:lag2} \\ 
    & \forall i= 1,\ldots d-1, \nonumber\\
    \frac{\partial\mathcal{L}}{\partial\lambda_{i,ki}} &= \log_2 \lambda_{i,ki} +\frac{1}{\ln2}+\mu_k+\mu_\text{N}=0 \label{eq:lag3} \\ 
    &\forall i = 1,\ldots,d-1,~k = 0,\ldots,m-2, \nonumber \\
    \frac{\partial\mathcal{L}}{\partial\lambda_{i,j}} &= \log_2 \lambda_{i,j} +\frac{1}{\ln2}+\mu_\text{N}=0\, \label{eq:lag4} \\
    &\forall i=1,\ldots,d-1,j\not\equiv ki~(\text{mod }d). \nonumber
\end{align}
We stress again that, if $d$ is not a prime, the equations above are only valid for $m\leq 3$.
Equating the constraints in Eq.~(\ref{eq:lag2}) for different $i$ yields directly $\lambda_{0i} = \lambda_{0j}$ or $\lambda_{0i} = 0$ for all $i,j$. It can be easily checked from  Eq.~(\ref{lagrangian}) that the choice $\lambda_{0i}=0$ only increases the key rate and therefore cannot be the minimum.
Likewise, $\lambda_{i,ki} = \lambda_{k,kj}$ for all $i,j$ and $k=1,\ldots,m-2$. From  
Eq.~(\ref{eq:lag4}), we obtain that those $\lambda_{i,j}$ which are not covered in Eqs.~(\ref{eq:lag1}) to (\ref{eq:lag3}) are also equal. From now on, we denote by $\lambda_Z$ any of the set of the equal coefficients $\{\lambda_{01},\ldots,\lambda_{0,d-1}\}$, by $\lambda_k$ any of the $\lambda_{i,ki}$ and by $\eta$ those coefficients which do not contribute to any of the error rates. The number of $\eta$ coefficients is $(d-1)^2 - (m-2)(d-1)$.

Now we need to distinguish two cases: $m=d+1$ and $m < d+1$. Indeed, in the first case our optimization provides and exact solution for each coefficient $\lambda_{\alpha,\beta}$ while, if $m < d+1$, the number of constraints is not large enough to fix all of them.

\subsection*{Case $m=d+1$}
If $m=d+1$, i.e., Alice and Bob measure a maximal set of MUBs, then Eq.~(\ref{eq:lag4}) becomes trivial, as each coefficient appears in exactly one of the error rate formulas. 

In this case, no additional optimization has to be performed and one can directly express the coefficients in terms of the observed error rates:
\begin{align}
    \lambda_Z &= \frac{q - Q_Z}{d-1}, \\
    \lambda_k &= \frac{q - Q_{XZ^k}}{d-1}, \\
    \lambda_{0,0} &=1- q,
\end{align}
with $q := (Q_Z + \sum_{k=0}^{d-1}Q_{XZ^k})/d$. Positivity requires that $Q_Z\leq q$, $Q_{XZ^k} \leq q$ for all $k$ and $q\leq 1$. By inserting these coefficients into the key rate formula, we find
\begin{align}\label{eq:aympt_asymm_dp1_app}
    r_\infty =& \log_2d+\lambda_{00}\log_2\lambda_{00} + (d-1)\lambda_Z \log_2\lambda_Z \nonumber \\
    &+(d-1)\sum_{k=0}^{d-1}\lambda_k \log_2 \lambda_k \nonumber \\
    =&\log_2d + (1-q)\log_2(1-q) + (q-Q_Z)\log(q-Q_Z) \nonumber \\
    &+ \sum_{k=0}^{d-1}(q-Q_{XZ^k})\log_2(q-Q_{XZ^k}) - q\log_2(d-1).
\end{align}
If all of the error rates are the same, i.e., $Q_Z=Q_{X}=...=Q_{XZ^{d-1}} \equiv Q$, then $q=(d+1)Q/d$ and the key rate simplifies to
\begin{align}\label{eq:aympt_symm_dp1_app}
    r_\infty^{(m=d+1)} = \log_2d -h(q) - q\log_2(d^2-1)
    \end{align}
with the binary Shannon entropy $h(q) = -q\log_2q - (1-q)\log_2(1-q)$.

\subsection*{Case $m<d+1$}
If $m<d+1$, the coefficient $\eta$ still has to be determined.
To that end, we consider the following linear combination of the constraints:
\begin{align}
    0&=\frac{\partial\mathcal{L}}{\partial\lambda_{0,0}} - \frac{\partial\mathcal{L}}{\partial\lambda_Z} - \sum_{k=0}^{m-2}\frac{\partial\mathcal{L}}{\partial\lambda_k} + (m-1)\frac{\partial\mathcal{L}}{\partial\eta} \nonumber \\
    &=\log_2\lambda_{0,0}+(m-1)\log_2\eta - \log_2(\lambda_Z\lambda_0\lambda_1\ldots\lambda_{m-2}),
\end{align}
which is equivalent to 
\begin{align}\label{eq:cond}
    \lambda_{0,0}\eta^{m-1} = \lambda_Z\lambda_0\lambda_1\ldots\lambda_{m-2}.
\end{align}
However, also the element $\lambda_{0,0}$ depends on $\eta$, as they are related via the normalization constraint
\begin{align}
    \lambda_{0,0} =& 1-(d-1)\lambda_Z - (d-1)\sum_{k=0}^{d-1}\lambda_k - [(d-1)^2 \nonumber \\
    &- (m-2)(d-1)]\eta.
\end{align}

Redefining $q = (Q_Z + \sum_{k=0}^{m-2}Q_{XZ^k})/(m-1)$, we obtain from the constraints
\begin{align}
    \lambda_Z &= \frac{q-Q_Z}{d-1} - \frac{d-(m-1)}{m-1}\eta,\\
    \lambda_k &= \frac{q-Q_{XZ^k}}{d-1} - \frac{d-(m-1)}{m-1}\eta,\\
    \lambda_{0,0} &= 1-q+(d-1)\frac{d-(m-1)}{m-1}\eta.
\end{align}
If we compress the prefactor of $\eta$ into $v:=(d-1)\frac{d-(m-1)}{m-1}$, we can rewrite the condition in Eq.~(\ref{eq:cond}) as
\begin{align}\label{eq:conditionlambda}
    &(d-1)^m (1-q+v\eta) \eta^{m-1} =\nonumber \\
    &(q-Q_Z-v\eta)(q-Q_X-v\eta)\ldots(q-Q_{XZ^{m-2}}-v\eta).
\end{align}
This yields a condition for the value of $\eta$, which in most cases cannot be expressed in a closed form. Inserting this relation into the key rate yields
\begin{align}\label{eq:asymp_lambda_app}
    r_\infty &= \log_2\frac{d}{d-1} - (m-1)(1-q)\log_2(\eta(d-1)) \nonumber \\
    &\phantom{=}+ (1-Q_Z)\log_2(q-Q_Z-v\eta) \nonumber\\
    &\phantom{=}+ \sum_{k=0}^{m-2}(1-Q_{XZ^k})\log_2(q-Q_{XZ^k}-v\eta).
\end{align}
The case of $m=2$ admits further simplification. Here, Eq.~(\ref{eq:conditionlambda}) simplifies to 
\begin{align}
    \eta = \frac{Q_Z Q_X}{(d-1)^2}.   
\end{align}
By inserting the explicit expressions for $\lambda_Z, \lambda_k$ and $\eta$ into the key rate, we obtain
\begin{align}\label{eq:aympt_asymm_2_app}
    r_\infty^{(m=2)} =& \log_2d - h(Q_X) - h(Q_Z) \nonumber \\ &- (Q_X+Q_Z)\log_2(d-1),
\end{align}
which, in the symmetric case $Q_X=Q_Z$ reduces to the asymptotic key rate for two MUBs as reported in Ref.~\cite{grasselli2020quantum}.

\bibliographystyle{apsrev4-1}
\bibliography{cite}

@article{bradler2016finite,
  title={Finite-key security analysis for multilevel quantum key distribution},
  author={Br{\'a}dler, Kamil and Mirhosseini, Mohammad and Fickler, Robert and Broadbent, Anne and Boyd, Robert},
  journal={New J. Phys.},
  volume={18},
  number={7},
  pages={073030},
  year={2016},
  publisher={IOP Publishing}
}

@phdthesis{grasselli2020quantum,
  title={Quantum Cryptography: from Key Distribution to Conference Key Agreement},
  author={Grasselli, Federico},
  year={2020},
  school={Universit\"at D\"usseldorf}
}

@article{wang2017generation,
  title={Generation of the complete four-dimensional Bell basis},
  author={Wang, Feiran and Erhard, Manuel and Babazadeh, Amin and Malik, Mehul and Krenn, Mario and Zeilinger, Anton},
  journal={Optica},
  volume={4},
  number={12},
  pages={1462--1467},
  year={2017},
  publisher={Optical Society of America}
}

@article{tomamichel2012tight,
  title={Tight finite-key analysis for quantum cryptography},
  author={Tomamichel, Marco and Lim, Charles Ci Wen and Gisin, Nicolas and Renner, Renato},
  journal={Nat. Commun.},
  volume={3},
  number={1},
  pages={634},
  year={2012},
  publisher={Nature Publishing Group}
}

@article{devetak2005distillation,
  title={Distillation of secret key and entanglement from quantum states},
  author={Devetak, Igor and Winter, Andreas},
  journal={Proceedings of the Royal Society A: Mathematical, Physical and engineering sciences},
  volume={461},
  number={2053},
  pages={207--235},
  year={2005},
  publisher={The Royal Society}
}

@article{tomamichel2011uncertainty,
  title={Uncertainty relation for smooth entropies},
  author={Tomamichel, Marco and Renner, Renato},
  journal={Phys. Rev. Lett.},
  volume={106},
  number={11},
  pages={110506},
  year={2011},
  publisher={APS}
}

@article{sheridan2010rapid,
  title={Security proof for quantum key distribution using qudit systems},
  author={Sheridan, Lana and Scarani, Valerio},
  journal={Phys. Rev. A},
  volume={82},
  number={3},
  pages={030301(R)},
  year={2010},
  publisher={APS}
}

@article{sheridan2010njp,
  title={Finite-key security against coherent attacks in quantum key distribution},
  author={Sheridan, Lana and Phuc Le, Thinh and Scarani, Valerio},
  journal={New J. Phys.},
  volume={12},
  number={},
  pages={123019},
  year={2010},
  publisher={IOP Publishing}
}

@article{renner2005,
  title={Information-theoretic security proof for quantum-key-distribution protocols},
  author={Renner, Renato and Gisin, Nicolas and Kraus, Barbara},
  journal={Phys. Rev. A},
  volume={72},
  number={},
  pages={012332},
  year={2005},
  publisher={APS}
}

@article{renner2008security,
  title={Security of quantum key distribution},
  author={Renner, Renato},
  journal={International Journal of Quantum Information},
  volume={6},
  number={01},
  pages={1},
  year={2008},
  publisher={World Scientific}
}

@article{renner2007symmetry,
  title={Symmetry of large physical systems implies independence of subsystems},
  author={Renner, Renato},
  journal={Nature Phys.},
  volume={3},
  number={},
  pages={645},
  year={2007},
  publisher={Nature Publishing Group}
}

@article{renner2009postselection,
  title={Postselection Technique for Quantum Channels with Applications to Quantum Cryptography},
  author={Christandl, Matthias and K\"onig, Robert and Renner, Renato},
  journal={Phys. Rev. Lett.},
  volume={102},
  number={2},
  pages={020504},
  year={2009},
  publisher={American Physical Society}
}

@article{durt2010,
  title={On mutually unbiased bases},
  author={Durt, Thomas and Englert, Berthold-Georg and Bengtsson, Ingemar and \.Zyczkowski},
  journal={Int. J. Quantum Inf.},
  volume={8},
  number={4},
  pages={535},
  year={2010},
  publisher={World Scientific}
}

@article{englert2001,
  title={The mean king's problem: prime degrees of freedom},
  author={Englert, Berthold-Georg and Aharonov, Yakir},
  journal={Phys. Lett. A},
  volume={284},
  number={},
  pages={1},
  year={2001},
  publisher={Elsevier}
}

@article{wootters1988,
  title={Optimal state-determination by mutually unbiased measurements},
  author={Wootters, William K. and Fields, Brian D.},
  journal={Ann. Phys.},
  volume={191},
  number={},
  pages={363},
  year={1988},
  publisher={Elsevier}
}

@article{tomamichel2010,
  title={Duality between smooth min- and max-entropies},
  author={Tomamichel, M. and Colbeck, R. and Renner, R.},
  journal={IEEE Trans. Inf. Theory},
  volume={54},
  number={},
  pages={4674},
  year={2010},
  publisher={IEEE}
}

@article{renner2005smoothent,
  title={Simple and Tight Bounds for Information Reconciliation and Privacy Amplification},
  author={Renner, R. and Wolf, S.},
  journal={Roy, B. (eds) Advances in Cryptology - ASIACRYPT 2005. ASIACRYPT 2005. Lecture Notes in Computer Science},
  volume={3788},
  number={},
  pages={199},
  year={2005},
  publisher={Springer}
}

@article{brecht2015,
  title={Photon Temporal Modes: A Complete Framework for Quantum Information Science},
  author={Brecht, B. and Reddy, Dileep V. and Silberhorn, C. and Raymer, M.~G.},
  journal={Phys. Rev. X},
  volume={5},
  number={},
  pages={041017},
  year={2015},
  publisher={American Physical Society}
}

@article{ali-khan2007,
  title={Large-Alphabet Quantum Key Distribution Using Energy-Time Entangled Bipartite States},
  author={Ali-Khan, I. and Broadbent, C. and Howell, J. and Raymer, M.~G.},
  journal={Phys. Rev. Lett.},
  volume={98},
  number={},
  pages={060503},
  year={2007},
  publisher={American Physical Society}
}

@article{walborn2006,
  title={Quantum key distribution with higher-order alphabets using spatially encoded qudits},
  author={Walborn, S. and Lemelle, D. and Almeida, M. and Ribeiro, P.},
  journal={Phys. Rev. Lett.},
  volume={96},
  number={},
  pages={090501},
  year={2006},
  publisher={American Physical Society}
}

@article{mirhosseini2015,
  title={High-dimensional quantum cryptography with twisted light},
  author={Mirhosseini, M. and Maga$\tilde{\text{n}}$a-Loaiza, O.~S. and O'Sullivan, M.~N. and Rodenburg, B. and Malik, M. and Lavery, M.~P.~J. and Padgett, M.~J. and Gauthier, D.~J. and Boyd, R.~W.},
  journal={New J. Phys.},
  volume={17},
  number={},
  pages={033033},
  year={2015},
  publisher={IOP Publishing}
}

@article{nussbaum2015,
  title={Toward frequency multiplexing for time-bin states},
  author={Nussbaum, B.~E. and Purakayastha, U. and Floyd, J. and Szuniewicz, J. and So\'snicki, F. and Karpi\'nski', M. and Kwiat, P.~G.},
  journal={Proc. SPIE 12633, Photonics for Quantum 2023},
  volume={12633},
  number={},
  pages={1263302},
  year={2023},
  publisher={SPIE}
}

@article{islam2017,
  title={Provably secure and high-rate quantum key
distribution with time-bin qudits},
  author={Islam, N.~T. and Ci Wen Lim, C. and Cahall, C. and Kim, J. and Gauthier, D.~J.},
  journal={Sci. Adv.},
  volume={3},
  number={},
  pages={e1701491},
  year={2017},
  publisher={American Association for the Advancement of Science}
}

@article{cerf2002,
  title={Security of Quantum Key Distribution Using d-Level Systems},
  author={Cerf, N.~J. and Bourennane, M. and Karlsson, A. and Gisin, N.},
  journal={Phys. Rev. Lett.},
  volume={88},
  number={},
  pages={127902},
  year={2002},
  publisher={The American Physical Society}
}

@article{bruss2002,
  title={Optimal Eavesdropping in Cryptography with Three-Dimensional Quantum States},
  author={Bru{\ss}, D. and Macchiavello, C.},
  journal={Phys. Rev. Lett.},
  volume={88},
  number={},
  pages={127901},
  year={2002},
  publisher={The American Physical Society}
}

@article{bruss1998,
  title={Optimal Eavesdropping in Quantum Cryptography with Six States},
  author={Bru{\ss}, D.},
  journal={Phys. Rev. Lett.},
  volume={81},
  number={},
  pages={3018},
  year={1998},
  publisher={The American Physical Society}
}

@article{biham1997,
  title={Security of Quantum Cryptography against Collective Attacks},
  author={Biham, E. and Mor, T.},
  journal={Phys. Rev. Lett.},
  volume={78},
  number={11},
  pages={2256},
  year={1997},
  publisher={The American Physical Society}
}

@article{scarani2009,
  title={The security of practical quantum key distribution},
  author={Scarani, V. and Bechmann-Pasquinucci, H. and Cerf, N.~J. and Du\v{s}ek, M. and L\"{u}tkenhaus, N. and Peev, M.},
  journal={Rev. Mod. Phys.},
  volume={81},
  number={},
  pages={1301},
  year={2009},
  publisher={The American Physical Society}
}

@article{ikuta2022,
  title={Scalable implementation of $(d+1)$ mutually unbiased bases for $d$-dimensional quantum key distribution},
  author={Ikuta, T. and Akibue, S. and Yonezu, Y. and Honjo, T. and Takesue, H. and Inoue, K.},
  journal={Phys. Rev. Res.},
  volume={4},
  number={},
  pages={L042007},
  year={2022},
  publisher={The American Physical Society}
}

@article{tomamichel2017,
  title={A largely self-contained and complete security proof for quantum key distribution},
  author={Tomamichel, M. and Leverrier, A.},
  journal={Quantum},
  volume={1},
  number={},
  pages={14},
  year={2017},
  publisher={Verein zur Förderung des Open Access Publizierens in den Quantenwissenschaften}
}

@article{mink2012,
  title={LDPC for QKD Reconciliation},
  author={Mink, A. and Nakassis, A.},
  journal={arXiv:1205.4977 [cs.CR]},
  volume={},
  number={},
  pages={},
  year={2012},
  publisher={}
}

@article{tomamichel2016,
  title={Quantum coding with finite resources},
  author={Tomamichel, M. and Berta, M. and Renes, J.~M.},
  journal={Nat. Commun.},
  volume={7},
  number={},
  pages={11419},
  year={2016},
  publisher={Nature Publishing Group}
}

@article{bennett1992,
  title={Quantum Cryptography without Bell's Theorem},
  author={Bennett, C.~H. and Brassard, G. and Mermin, N.~D.},
  journal={Phys. Rev. Lett.},
  volume={68},
  number={},
  pages={557},
  year={1992},
  publisher={The American Physical Society}
}

@article{wang2021,
  title={Tight finite-key analysis for generalized high-dimensional quantum key distribution},
  author={Wang, R. and Yin, Z.-Q. and Liu, H. and Wang, S. and Chen, W. and Guo, C.-G. and Han, Z.~F.},
  journal={Phys. Rev. Res.},
  volume={3},
  number={},
  pages={023019},
  year={2021},
  publisher={The American Physical Society}
}

@article{bb84,
  title={Quantum cryptography: Public key distribution and coin tossing},
  author={Bennett, C.~H. and Brassard, G.},
  journal={Proceedings of IEEE International Conference on Computers, Systems and Signal Processing},
  volume={175},
  number={},
  pages={8},
  year={1984},
  publisher={}
}

@article{doda2021,
  title={Quantum key distribution overcoming extreme noise: simultaneous subspace coding using high-dimensional entanglement},
  author={Doda, M. and Huber, M. and Murta, G. and Pivoluska, M. and Plesch, M. and Vlachou, C.},
  journal={Phys. Rev. Appl.},
  volume={15},
  number={},
  pages={034003},
  year={2021},
  publisher={American Physical Society}
}

@article{araujo2023,
  title={Quantum key distribution rates from semidefinite programming},
  author={Ara\'{u}jo, M. and Huber, M. and Navascu\'{e}s, M. and Pivoluska, M. and Tavakoli, A.},
  journal={Quantum},
  volume={7},
  number={},
  pages={1019},
  year={2023},
  publisher={}
}

@article{winick2018,
  title={Reliable numerical key rates for quantum key distribution},
  author={Winick, A. and L\"{u}tkenhaus, N. and Coles, P.~J.},
  journal={Quantum},
  volume={2},
  number={},
  pages={77},
  year={2018},
  publisher={}
}

@article{brown2021,
  title={Computing conditional entropies for quantum correlations},
  author={Brown, P. and Fawzi, H. and Fawzi, O.},
  journal={Nat. Commun.},
  volume={12},
  number={},
  pages={575},
  year={2021},
  publisher={}
}

@article{Hu2022,
  title={Robust Interior Point Method for Quantum Key Distribution Rate Computation},
  author={Hu, H. and Im, J. and Lin, J. and L\"{u}tkenhaus, N. and Wolkowicz, H.},
  journal={Quantum},
  volume={6},
  number={},
  pages={792},
  year={2022},
  publisher={}
}

@article{Lib2024,
  title={High-dimensional quantum key distribution using a multi-plane light converter},
  author={Lib, O. and Sulimany, K. and Lin, J. and Ara\'{u}jo, M. and Ben-Or, M. and Bromberg, Y.},
  journal={Optica Quantum},
  volume={3},
  pages={182},
  year={2025}
}

@article{Zahidy2024,
  title={Practical high-dimensional quantum key distribution protocol over deployed multicore fiber},
  author={Zahidy, M. and Ribezzo, D. and De Lazzari, C. and Vagniluca, I. and Biagi, N. and M\"{u}ller, R. and Occhipinti, T. and Oxenl{\o}we L.~K. and Galili, M. and Hayashi, T. and Cassioli, D. and Mecozzi, A. and Antonelli, C. and Zavatta, A. and Bacco, D.},
  journal={Nat. Commun.},
  volume={15},
  number={},
  pages={1651},
  year={2024},
  publisher={}
}

@article{Sekga2023,
  title={High-dimensional quantum key distribution implemented with biphotons},
  author={Sekga, C. and Mafu, M. and Senekane, M.},
  journal={Sci. Rep.},
  volume={13},
  number={},
  pages={1229},
  year={2023},
  publisher={}
}

@article{canas2017,
  title={High-dimensional decoy-state quantum key distribution over multicore telecommunication fibers},
  author={Ca\~nas, G. and Vera, N. and Cari\~ne, J. and Gonz\'alez, P. and Cardenas, J. and Connolly, P. W. R. and Przysiezna, A. and G\'omez, E. S. and Figueroa, M. and Vallone, G. and Villoresi, P. and da Silva, T. Ferreira and Xavier, G. B. and Lima, G.},
  journal={Phys. Rev. A},
  volume={96},
  number={},
  pages={022317},
  year={2017},
  publisher={IOP Publishing}
}

@book{tomamichelbook,
    author = {Tomamichel, M.},
    title = {Quantum Information Processing with Finite Resources},
    publisher = {Springer},
    year = {2016} 
}

@article{serfling1974,
  title={Probability Inequalities for the Sum in Sampling without Replacement},
  author={Serfling, R.~J.},
  journal={Ann. Statist.},
  volume={2},
  number={},
  pages={39},
  year={1974},
  publisher={Institute of Mathematical Statistics}
}

@ARTICLE{tomamichel2009,
  author={Tomamichel, Marco and Colbeck, Roger and Renner, Renato},
  journal={IEEE Trans. Inf. Theory}, 
  title={A Fully Quantum Asymptotic Equipartition Property}, 
  year={2009},
  volume={55},
  number={12},
  pages={5840-5847},
  doi={10.1109/TIT.2009.2032797}}

@ARTICLE{huber2024,
  author={Kanitschar, Florian and Huber, Marcus},
  journal={Phys. Rev. Lett.}, 
  title={A practical framework for analyzing high-dimensional QKD setups}, 
  year={2025},
  volume={135},
  pages={010802}
  }

@article{gisin1999,
  title = {Incoherent and coherent eavesdropping in the six-state protocol of quantum cryptography},
  author = {Bechmann-Pasquinucci, H. and Gisin, N.},
  journal = {Phys. Rev. A},
  volume = {59},
  issue = {6},
  pages = {4238--4248},
  numpages = {0},
  year = {1999},
  month = {Jun},
  publisher = {American Physical Society},
  doi = {10.1103/PhysRevA.59.4238},
  url = {https://link.aps.org/doi/10.1103/PhysRevA.59.4238}
}

@article{nahar2024postselection,
  title={Postselection technique for optical Quantum Key Distribution with improved de Finetti reductions},
  author={Nahar, Shlok and Tupkary, Devashish and Zhao, Yuming and L{\"u}tkenhaus, Norbert and Tan, Ernest Y-Z},
  journal={PRX Quantum},
  volume={5},
  number={4},
  pages={040315},
  year={2024},
  publisher={APS}
}

\end{document}